\documentclass[pre,twocolumn,showpacs,preprintnumbers,amsmath,amssymb,floatfix]{revtex4}
\def\beneq{\begin{equation}}
\def\eneq{\end{equation}}
\def\bea{\begin{eqnarray}}
\def\eea{\end{eqnarray}}
\usepackage{graphicx}% Include figure files
\usepackage{dcolumn}% Align table columns on decimal point
\usepackage{bm}% bold math
\begin{document}

%%%%%%%%%%%%%%%%%%%%%
%  Version 8
% Last Changes made by AJL
% On November 27, 2007
%%%%%%%%%%%%%%%%%%%%%%
\title{Rheology of two-dimensional F-actin networks associated with a lipid interface}

\author{Robert Walder,$^{1}$, Alex J. Levine$^{2,3}$, and Michael Dennin$^{1}$}

\affiliation{
  $^{1}$Department of Physics \& Astronomy, University of California, Irvine, CA 92697\\
  $^{2}$Department of Chemistry \& Biochemistry, University of California, Los Angeles, CA 90095\\
  $^{3}$California Nanosystems Institute, UCLA, Los Angeles, CA
  90095
}

\date{\today}

\begin{abstract}

  We report on the surface rheology of cross-linked F-actin networks
  associated with a lipid monolayer at the air-water interface of a
  Langmuir monolayer. The rheological measurements are made using a
  Couette cell. These data demonstrate that the network has a finite
  elastic modulus that grows as a function of the cross-linking
  concentration. We also note that under steady-state flow the
  system behaves as a power law fluid in which the effective
  viscosity decreases with imposed shear.
\end{abstract}

\pacs{83.60.Df,% Nonlinear viscoelasticity
68.08.-p,% Liquid-solid interfaces
82.35.Pq %Biopolymers, bio-polymerization
82.70.Gg,% Gels and sols
}

\maketitle

\section{Introduction}

The cytoskeleton of eukaryotic cells is a chemically heterogeneous
assemblage of filamentous proteins, cross-linking agents, and
molecular motors. This stiff filamentous network is mechanically
linked to the plasma membrane and comprises
the principal structural element of the cell. It confers mechanical
stability and is the site of force generation and morphological
control of the cell body. As such it is essential for such biologically
important processes as intracellular trafficking, cell motility, and
the measurement of applied stress.

In the past great progress has been made in elucidating the
underlying physical properties of the cytoskeleton by building
highly simplified {\em in vitro} model systems, which provide
controlled laboratories for exploring the mechanics of chemically
well-controlled systems extracted from the complex cytoskeleton.
From this work has emerged a well-developed model of semiflexible
F-actin networks and solutions. While much is currently known
about the mechanics of semiflexible
networks~\cite{Janmey:90,MacKintosh:97,Gardel:04,Gardel:06}, less
is understood about the mechanical properties of this biopolymer
network when mechanically associated with the fluid plasma
membrane of the cell. One reason for this is that the coupling of
a semiflexible network to the cell membrane is a distinct feature
of cellular
mechanics~\cite{Desprat:05,Coughlin:06,VanCitters:06,Trepat:07}
that has not been as well studied using simplified {\em in vitro}
models.  Previous work in this area has been done using
indentation-based mechanical probes of F-actin coated lipid
vesicles~\cite{Helfer:01} and
microrheology~\cite{Helfer:00,Helfer:01a}.

% made new paragraph and expanded on the linker issue of referee A

In this work we present a new biomimetic model for this composite
structure that we believe is more amenable to detailed mechanical
study and the visualization of the strain field within it. We have
developed a cross-linked F-actin network that is mechanically
associated with a DPPC monolayer in a Langmuir trough by
biotin/streptavidin bonds. This coupling approximates the more
complex linkage between the physiological cytoskeleton and the
plasma membrane effected via specialized protein
complexes~\cite{Brunton:04}. It is important to keep in mind that
this is truly a minimal model of the composite
membrane/cytoskeleton system.  The full biological system
includes significantly more complex linkages between actin
filaments and the actin-lipid connection. For example, there are
multiple inter-filament cross-links, some of which introduce
interesting elastic properties (such as Filamin linkers)~\cite{DiDonna:06,DiDonna:07,Broederaz:07}.
Many physiologically relevant cross-linking proteins are labile generating
transient bonds.  The cross-linking scheme employed here, however,
creates permanent cross-links on the time-scale of the experiments. We
briefly speculate on the rheological signature of such labile cross-linkers
in the summary of this work.
Even though the choice of a simpler system of linkage molecules
severely limits its chemical complexity, it is precisely this
choice of system that allows for reproducible, quantitative
measurements in a system that is more amenable to theoretical
investigation~\cite{MacKintosh:95,Head:03,Wilhelm:03}.

%paragraph added to address referee B concerns
We suspect that the rheological effect of the monolayer is
indirect. In this work, the lipid monolayer is in a highly fluid
phase, so that it has no zero-frequency elastic modulus. We
estimate that, at the frequencies probed here, the viscous
stresses in the lipid monolayer are 2 - 3 orders of magnitude less
than the elastic stresses in the actin network. Therefore, we do
not expect it to make any direct contribution to the measured
mechanical properties of the complex. By creating a thin-layer of
F-actin bound to the monolayer, however, we anticipate that we
have likely generated an anisotropic network in which the long
F-actin filaments are tangent to the lipid decorated surface. This
anisotropic structure reflects the coarse geometry of the cell's
dense cortical cytoskeleton, and we explore the mechanics of this
highly anisotropic network. Additionally, the monolayer serves as
a protective barrier for the actin, preventing denaturation at the
air-water interface. The thin-film geometry has other benefits not
fully exploited in this work: one can directly examine the
resulting planar strain field under stress. This feature of the
system will be more fully discussed elsewhere. Finally, one can
consider the binding of the network to denser lipid monolayers
whose elastic moduli can compete with those of the F-actin
network. However, the impact of denser monolayers on the response
of the system will be the subject of future
work~\cite{Walder:07a}.

We rheologically probe this network using a newly developed
Couette-style, surface rheometer, the details of which will be
presented elsewhere~\cite{Walder:07}. In this first report on the
mechanics of the F-actin/monolayer system, we focus primarily on
the steady-state shearing of the network. In this state we expect
that the F-actin network flows due to continual rearrangements of
cross-linked or sterically jammed collections of filaments. These
filament networks must transiently interpenetrate each other and
then disengage during the flow. Though this regime does not have
direct biological applications, it is important for the complete
material characterization of the system. It provides important
baseline measurements on the bulk elastic and viscous properties
that will be useful in future microrheological studies of the
network.

Under steady-state shear, we find that the material is highly
nonlinear in its rheological response behaving as a
shear-thinning, power-law fluid.  The power-law exponent is
weakly dependent on the density of permanent cross-links in the
system. We find no evidence for a yield stress, and the observed
elastic modulus is weakly dependent on shear rate.  Based on these
data we believe that our network consists of permanently bonded
patches of filaments that interpenetrate each other; these patches
exchange nearest neighbors during the shear. As such the network
may be characterized as having many metastable states separated by
low energy barriers. It has been suggested that such system should
obey a type of universal ``soft glassy rheology.'' We find that
our rheological data in this steadily flowing state is indeed
consistent with the predictions of this theory~\cite{Sollich:97}.

By examining the initiation and cessation of shear, we explore the
time-dependent viscoelastic response of the  material. We find
that features of the linear response of the F-actin network can be
accounted for using a generalized Maxwell model having two stress
relaxation times. Our data suggest that the elastic modulus of the
network grows weakly, but in a linear fashion with cross-linker density;  it
also plateaus at higher cross-linking densities suggesting that steric hinderances
prevent the saturation of all available cross-linkers in the
system.

In the remainder of this article, we first discuss briefly the
experimental design of the apparatus, as well our methods for
analyzing the data in section \ref{Materials-Methods}. We then
present our data and discuss its interpretation in section
\ref{results},  before concluding with a discussion of the
implications of this work to future investigations in section
\ref{conclusions}.

\section{Materials and methods}
\label{Materials-Methods}
\subsection{Preparation of actin-lipid composite material}

We create the F-actin monolayer {\em in vitro} by polymerizing for
one hour G-actin monomers along with the variable fraction of
biotynlated actin monomers in a F-actin buffer (Tris HCl buffer
pH=7.4, 0.2 mM ATP, 3 mM MgCl$_2$, 0.2 mM CaCl$_{2}$). In each
preparation we fixed the concentration of the non-biotynlated
actin to 2 mg/mL. We use the parameter R, which is the ratio of
biotynlated actin monomers to G-actin monomers, to characterize
the cross linking density of our actin filament networks. We
polymerize the actin using this standard procedure to ensure that
actin filaments are formed. With this procedure, we expect
filaments that average $5~\ {\rm \mu m}$ in length. Because our
apparatus requires relatively large volumes of fluid, we diluted
these actin solutions 100X and added Rhodamine phalloidin and
streptavidin. The phalloidin
stabilizes the F-actin filaments against depolymerization during
the experiment~\cite{Wehland:77} while the streptavidin
cross-links the biotynlated monomers creating a permanent,
stress-bearing semiflexible polymer network.
Once diluted, we estimate that the mesh size of the
F-actin network is $2.7\mu$m~\cite{Schmidt:89}.  We waited $30$
minutes for the completion of the cross-linking reaction before
transferring the protein solution into the Couette trough and
adding a lipid solution (93\% DPPC, 7\% DPPE - Biotynl) at the
surface. We then allowed the system to equilibrate for one hour in
order for the F-actin network to diffuse to and irreversibly bind
to the biotynlated monolayer~\cite{Dichtl:99}. We then performed
the rheological measurements over a period of about two hours.
During this period we observed no evidence of a significant change
in rheological properties reflecting either continued
cross-linking or network degradation. These measurements were made
using a Couette surface viscometer designed specifically for this
work.

%From stoichiometric considerations using the assumption that all
%F-actin filaments become irreversibly bound to the monolayer,  and the area
%density of the filaments is approximately $1700/ \mu \mbox{m}^2$
%at the surface. Furthermore, based on the density of biotynlated
%actin monomers, we estimate that there are approximately $10-80$
%cross-links per filament. {\bf Really don't think this is right.}

The depth profile of actin-lipid composite materials formed at the
air-water interface has been previously characterized by neutron
reflectivity experiments~\cite{Sackmann:06}. These experiments show
that that a monolayer of close packed actin filaments will bind to a
layer of streptavidin and thus irreversibly bind to the lipid monolayer.
Therefore, we expect that the we have created a low yield stress solid
associated with the air-water interface due to a combination of
biotyn--streptavidin cross-links and steric interactions between
filaments. In this set of rheological measurements, we examine
primarily the steady-state flow properties of this network under
time-independent shearing imposed at the boundaries. We will explore
the small strain elastic behavior of this material in future
work~\cite{Walder:07a}.

\subsection{Couette Surface Rheology with Torsion Pendulum}

\begin{figure}
\includegraphics[width=8cm]{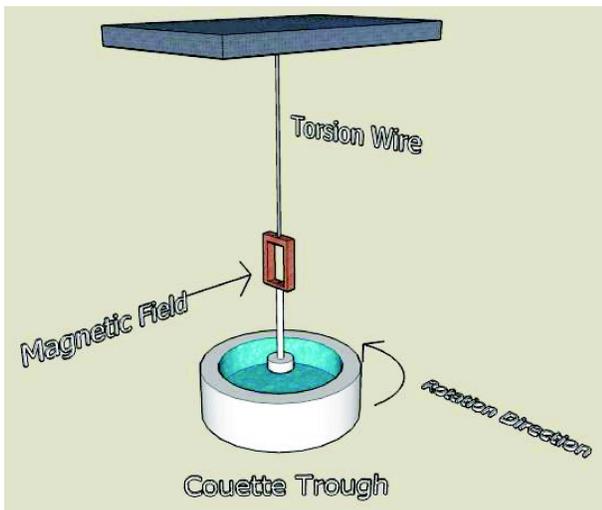}
\caption{A schematic diagram of the Couette trough used in the
experiments. The F-actin and lipid monolayer completely fills the
surface between the inner and outer teflon cylinders. The outer
cylinder is rotated at constant angular velocity and angular
displacement of the inner cylinder is monitored via the magnet
rigidly coupled to torsion wire.} \label{CouetteTrough}
\end{figure}

The Couette surface viscometer is a modified Couette rheometer
(see~\cite{Dennin:98} for a description of the main principles of
operation) designed specifically to measure the viscosity of
surface films. The details of the instrument will be described in
a separate publication~\cite{Walder:07}; however, its main
features are reviewed here. The apparatus consists of a teflon
cylindrical container forming the outer barrier (of radius $r_o =
3.75\ {\rm cm}$) of the sample and an teflon inner barrier (of
radius $r_i = 1.25\ {\rm cm}$) coupled to a torsion pendulum. A
wire coil attached to the torsion pendulum is positioned within a
second set of coils. An ac current in the second set of coils
generates a high frequency magnetic field that induces a voltage
in the coil attached to the torsion pendulum. The angular position
of the torsion pendulum is detected by the magnitude of the
induced voltage in the attached coil. A schematic diagram of the
experiment is shown in Fig.~\ref{CouetteTrough}.

We used a lock-in amplifier to detect the small induced voltages
at the same frequency but $90^\circ$ out of phase with the
reference signal powering the external magnetic field. The angular
position as a function of induced voltage is calibrated using an
angular translation stage. From the angular displacement we
calculate the torque ${\cal T}$ on the inner barrier using the
torsion constant of the torsion pendulum, which we determined
previously. The torque ${\cal T}$ measured at the inner rotor
(but constant throughout the system)
of radius $r_i$ can then be related to the shear stress there by
geometry:
\begin{equation}
\label{Torque-Stress}
\sigma_{r \theta}|_{r=r_i} = \frac{{\cal T}}{2 \pi r_i^2},
\end{equation}
where $r$ and $\theta$ are the radial and azimuthal coordinates
respectively on the surface of the Langmuir monolayer. We suppress
these indices hereafter. The raw data consists of torque versus
time curves under constant rate of strain or after the cessation
of applied strain. This data is converted into elastic modulus,
viscosity, and relaxation time constants using well-defined
procedures. However, two aspects of this rheometer are
non-standard: it operates in a wide-gap geometry and the inner
cylinder is free to move. Therefore, it is worth summarizing the
impact of these features on the measurements of the elastic modulus and the
viscosity. These will be discussed in the next section in the
context of an example torque versus time curve.

\section{Results}
\label{results}

As previously discussed, we focus on steady-state, constant shear
rate rheological studies to examine the changes in the rheology of
the network as a function of cross-linker density. We also study
the response of the system at the initiation and cessation of
shear to examine the linear, time-dependent shear modulus. The
studies of the initiation of flow allow for determination of a
linear elastic modulus for the system, while examining the stress
relaxation at the cessation of shear can be used to measure the
stress relaxation time in the medium. We parameterize the linear,
time-dependent stress response during the initiation and cessation
of flow by introducing a simple Maxwell fluid
model~\cite{Landau:86}. Examining the steady-state rheology, we
observe a strong shear-rate dependence of our results that is
well-described by a power-law fluid model~\cite{Bird:77}. Due to
this nonlinear behavior, we do not expect that the simple Maxwell
model will also capture the steady-state flow. By varying the
density of cross-linkers, we study how the nonlinearity of the
material under steady-state shear and how the elastic modulus of
the network changes due to changes in the density of
cross-linkers.

\begin{figure}[htb]
\includegraphics[width=6cm]{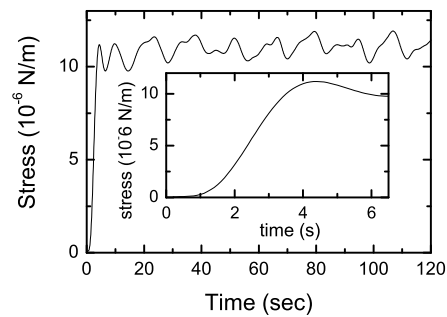}
\caption{Typical data set: The F-actin network has a cross linker
density of $R=0.005$. At $t=0$ the outer cylinder reaches a fixed
angular velocity of $0.4$ rad/sec (which corresponds to a shear
rate of $1.3\ {\rm s^{-1}}$). The measured torque rapidly reaches
a plateau and returns to zero when the outer cylinder is stopped
(see Fig.~\ref{fig:decayexample}). Insert shows a closeup of the
initial rise.} \label{fig:TorquevsTime}
\end{figure}

\begin{figure}[htb]
\includegraphics[width=6cm]{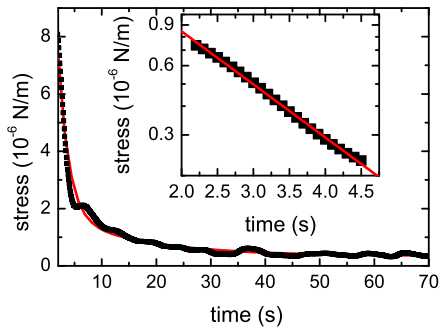}
\caption{(color online) The torque versus time for the cessation
of flow for the F-actin network with a cross linker density of
$R=0.005$. The outer cylinder was rotated with a fixed angular
velocity of $0.4$ rad/sec (shear rate 1.3), and stopped at t = 0
s. The solid line is a fit to a double exponential decay as
discussed in the text. Insert: Semi-log plot of the same data
showing just the early time behavior and the dominant relaxation
time. The solid red line is a fit to the data with time constant
1.9 s.} \label{fig:decayexample}
\end{figure}

All the data from the experiments performed using our Couette
style apparatus share similar phenomenology. We show a typical
example of the time evolution of the measured torque from the
initiation of shear to steady-state in
Fig.~\ref{fig:TorquevsTime}. In Fig.~\ref{fig:decayexample}, we
show stress decay at the cessation of shear  and a linear
fit to the essentially exponential decay of stress on a semilog
plot (inset). At times before $t=0$, the outer barrier is at rest; these
data are used to zero the measurement of the torque. At $t=0$, the
outer barrier begins to rotate at a fixed angular velocity. Given
the temporal resolution of the instrument we may consider this
change in angular velocity to be effectively instantaneous. After
a brief interval of stress overshoot, the system relaxes to a
steady-state torque, which can be interpreted in terms of the
steady-state viscous response of the surface film. The observed
fluctuations in the torque at steady-state system are larger in
magnitude than those found in the stationary system. Therefore,
the torque fluctuations in the steadily sheared state suggest the
quasi-periodic build up and decay of internal stress-bearing
structures associated with the monolayer. In this article,
however, we focus on the time-averaged response of the material in
the steady-state. Finally, once the rotation of the outer barrier
ceases, the measured torque is observed to decay to zero in an
exponential fashion -- see Fig.~\ref{fig:decayexample}. We first
describe how each section of the stress response is analyzed, and
then discuss the behavior as a function of cross-linker density.

The growth of the shear stress immediately after the initiation of
strain is used to extract information about the elastic modulus of
the interface (see insert in Fig.~\ref{fig:TorquevsTime}). In this
regime of small strains, the linear response of the stress is
assured so we may write~\cite{Doi:86}
\begin{equation}
\label{linear-response}
\sigma(t) =  \int_{0}^t dt' G(t-t') \gamma(t'),
\end{equation}
where $\gamma(t)$ is the strain history of the sample. We have
implicitly assumed that there was no strain before $t=0$.
Expressing this in the frequency domain, prescribing the strain to
be $\gamma(t) = \dot{\gamma} t \theta(t) $ where $\dot{\gamma}$ is
a constant, and $\theta(t)$ is the usual step function, and
expanding for $t \longrightarrow 0^{+}$, we may write
\begin{equation}
\label{short-times}
\sigma(t) = \dot{\gamma} \int_0^t dt' \int_{-\infty}^{\infty}
\frac{d \omega}{2 \pi} G^{*}(\omega) e^{i \omega (t -t')}
t' =  G'(0)  \dot{\gamma} t + {\cal O}(t^2),
\end{equation}
where $G^*(\omega) = G'(\omega) + i G''(\omega)$ is the Fourier
transform of the viscoelastic shear modulus appearing in
Eq.~\ref{linear-response}. Thus, for an ideal apparatus, we may
interpret the slope of the initial rise in the stress in terms of
the integral of the time-dependent elastic modulus
\begin{equation}
\label{elastic-mod-integral}
G'(0) = \lim_{t->0^{+}} \int_{-\infty}^\infty G'(\omega) e^{- i \omega t}  \frac{d \omega}{2 \pi}.
\end{equation}
If the storage modulus is dominated by its long-time or
low-frequency behavior, we may interpret the initial growth in the
stress in terms of the elastic modulus of F-actin network
associated with the Langmuir monolayer.

It should be noted that because the inner cylinder moves during
the initial shear, one has to define the initial strain carefully.
Therefore, in practice, the elastic modulus $G'(0) \equiv G$ is
measured using the following relation \cite{Dennin:04}:

\begin{equation}
\label{elastic_mod} G = \frac{\omega}{\Omega - \omega}\left(
\frac{\kappa}{4\pi}\right) \left(\frac{1}{r_i^2} - \frac{1}{r_o^2}
\right),
\end{equation}
where $\Omega$ is the angular rotation speed of the outer
cylinder, $\omega$ is the angular rotation speed of the inner
cylinder, $\kappa$ is the torsion constant of the supporting wire,
and $r_o$ is the outer cylinder radius. For more details on the
this procedure, see Ref.~\cite{Dennin:04}.

Because the initial stress rise is approximately linear and the
subsequent decay is exponential, it is reasonable to treat the
material as a simple Maxwell fluid having a single stress
relaxation time $\tau$, so that we may write
\begin{equation}
\label{Maxwell-Frequency-Domain}
G^{*}(\omega) = \frac{ G_0}{1 - i \omega \tau}
\end{equation}
for the viscoelastic shear modulus, then using
Eqs.~\ref{short-times} and \ref{elastic-mod-integral}, this allows
us to identify our previously defined elastic modulus (based on
the initial slope of the stress versus time curve) with $G_0$.

The other element of the Maxwell model is the stress relaxation
time $\tau$. At the cessation of shear (at time $T$), the stress
relaxes as
\begin{equation}
\label{Maxwell-stress-relaxation} \sigma(t) =  G_0 \dot{\gamma}
\tau  e^{-(t-T)/\tau},
\end{equation}
for times $t > T$. An example of the stress relaxation is given in
Fig.~\ref{fig:decayexample}. By considering the log of the stress
versus time, we are able to extract a relaxation time. Over the
range of cross-linker densities studied, we observed no dependence
of the relaxation time on cross-liner density. The value is $\tau
= (1.58 \pm 0.3)\ {\rm s}$. Closer examination of the stress
relaxation reveals that it is better fit by a double exponential
decay and that there is no significant improvement in the fit
resulting from the use of additional exponential decays. There is
a short relaxation time on the order of seconds for which the
exponential has an amplitude roughly an order of magnitude larger
than the second exponential. For the example in
Fig.~\ref{fig:decayexample}, the short relaxation time is 1.9 s
with amplitude $32.1 \times 10^{-6}\ {\rm N/m}$ and the long
relaxation time is 15.4 s with an amplitude of $1.8 \times
10^{-6}\ {\rm N/m}$.

In the context of the Maxwell model, the steady-state viscosity is
given by $\tau G_0$. This provides a consistency check on the
model. In our case,  if we use the value of $G_0$ from the flow
start-up, the values of $\tau G_0$ range between $10^{-3}$ and
$10^{-4}\ {\rm Ns/m}$ over the shear rates studied. In contrast,
the viscosity ranged from $10^{-4}$ and $10^{-5}\ {\rm Ns/m}$,
roughly a factor of 10 smaller. Surprisingly, both quantities had
roughly the same scaling with shear-rate. Combining this
particular break-down of the Maxwell model with the fact that the
relaxation times $\tau$ are independent of cross-linker density
suggests the following picture for the system. During the initial
applied strain, the system is sufficiently linear that we are able
to measure the elastic response due to both stretching of
cross-linked filaments and the interactions between domains of
cross-linked filaments. Once the system flows, the dissipation is
dominated by the interaction between the domains. Likewise, the
dominant stress relaxation is between domains, and therefore,
$\tau$ does not depend on the cross-linking concentration.
Finally, the break up of the domains during steady shear explains
the inconsistency between $G_0$ measured at start-up and from flow
cessation. It is the role played by the cross-linkers that results
in a larger measured elastic modulus than one would expect for the
Maxwell model.

\begin{figure}[htb]
\includegraphics[width=6cm]{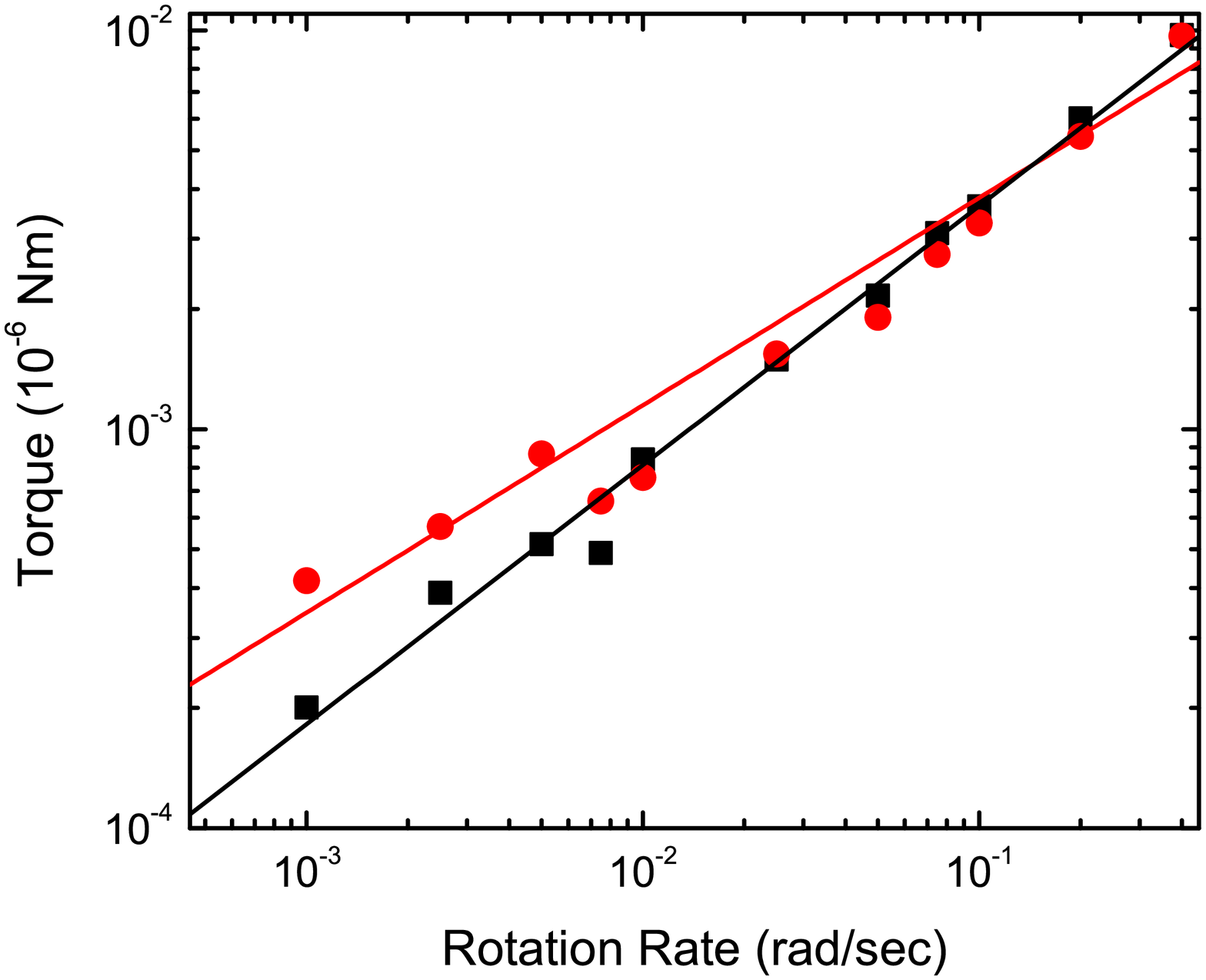}
\includegraphics[width=6cm]{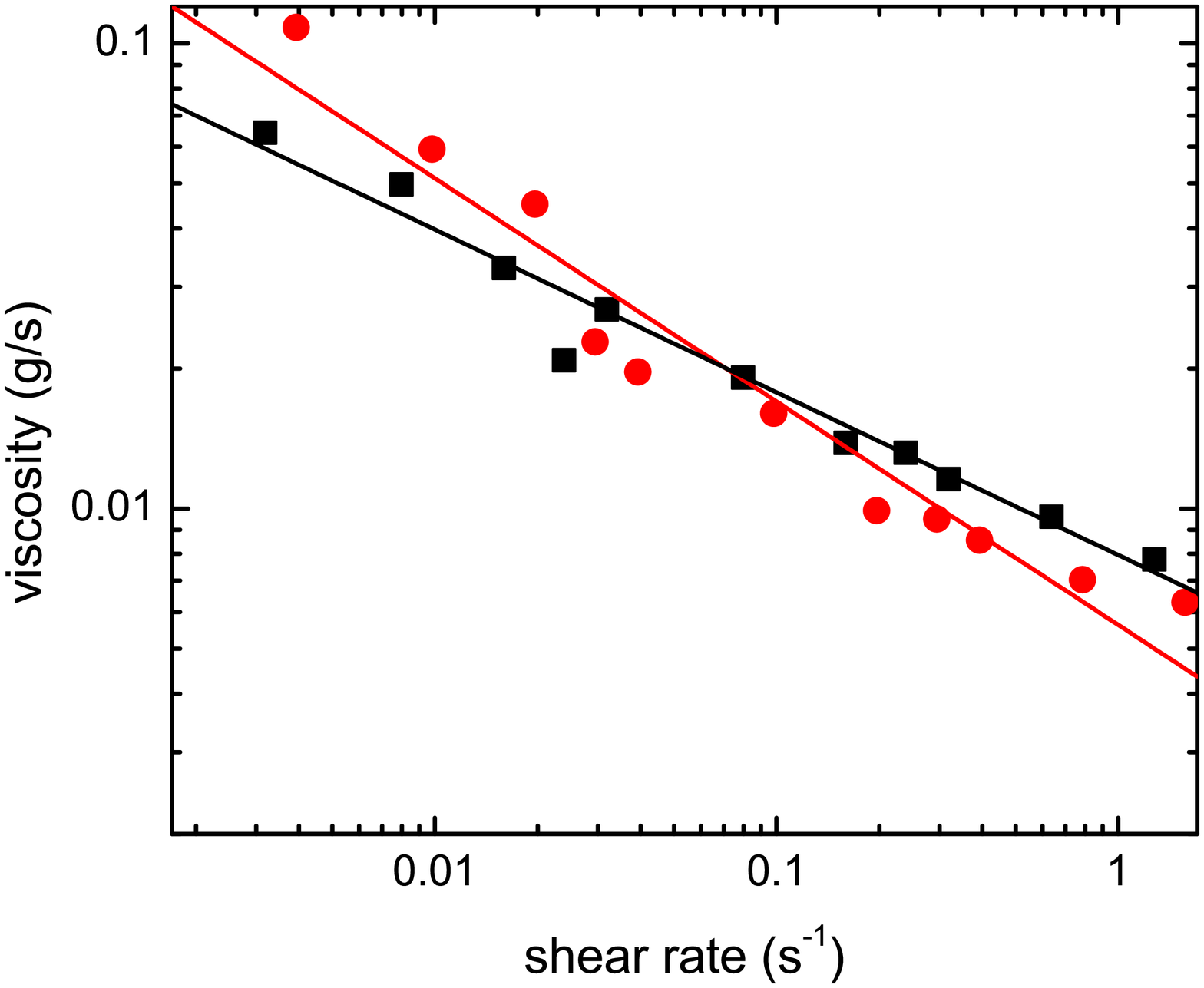}
\caption{(color online) Torque and surface viscosity measurements
on an uncross-linked F-actin network ($R=0$, black squares) and a
F-actin network with $R=0.04$ cross linker density (red circles)
at standard concentration associated with the lipid monolayer. (a)
Torque vs.\ rotation rate of the outer cylinder showing a
power-law dependence expected for a fluid in which the viscosity
decays as a power of shear rate. See Eq.~\ref{power-law-fluid}.
The lines are the best fits to the data. (b) The surface viscosity
of the composite monolayer extracted from the data in (a). See
Eqs.~\protect{\ref{power-viscosity-I},\ref{power-viscosity-II}}.
The lines are the best fit to the data.} \label{fig:power-law}
\end{figure}

To further test this picture, we consider the steady-state
viscosity in more detail. The steady-state viscosity of the system
is strongly shear-rate dependent (see Fig.~\ref{fig:power-law}).
This is another reason that we are not surprised at the failure of
the Maxwell model to completely describe the behavior, as the
Maxwell model is a linear response model. We return to this point
in our conclusions -- see section~\ref{conclusions}. It is worth
pointing out another sign of the failure of linear rheological
models; the observed stress overshoot at the initiation of flow --
see Figs.~\ref{fig:TorquevsTime}. Stress overshoot has been
observed in experiments~\cite{Raible:79,Osaki:00} and in molecular
dynamics simulations~\cite{Moore:99} of flexible polymer melts and
solutions although care must be taken to distinguish this effect
from systematic errors introduced by the
rheometer~\cite{Meissner:72}. Theoretical models of this nonlinear
rheological effect have been proposed for polymeric
liquids~\cite{Murayama:80,Pearson:91} and even metallic
glasses~\cite{Chen:00}.

Because our system operates in a wide-gap mode, we will review how
the viscosity is measured. Before discussing the nonlinear
response of our system, it is useful to recall the response of a
Newtonian fluid in a Couette rheometer with an arbitrary gap
width. Solving the Navier-Stokes equations for the Couette
flow~\cite{Landau:04} in steady-state between a rotating outer
cylinder of radius $r_o$ at speed $v_0 = \Omega r_o$ and a
stationary inner cylinder of radius $r_i$ we find that the only
non-vanishing component of the rate of strain tensor
$\dot{\gamma}_{r \theta} = \dot{\gamma}$ takes the form
\begin{equation}
\label{Couette-strain-rate} \dot{\gamma} = \frac{2 \Omega
r_o^2}{r_o^2 - r_i^2} \left(\frac{r_i}{r} \right)^2.
\end{equation}
For a viscous, Newtonian fluid with linear relationship between
the shear stress and the shear rate of strain, we find that the
torque can be related to the viscosity and rotation rate of the
outer cylinder by~\cite{Bird:77}
\begin{equation}
\label{torque-viscosity} {\cal T} =\frac{4\pi\eta r_i^2 r_o^2
\Omega}{(r_o^2-r_i^2)},
\end{equation}
where, $\eta$ is the surface viscosity.  The aqueous subphase
makes only a small contribution to the shear stresses in the
system, so using the above relation Eq.~\ref{torque-viscosity} we
can determine the effective viscosity of the interface in
steady-state. An important feature of this relationship is the
fact that the torque is a linear function of the rotation rate. In
Fig.~\ref{fig:power-law}(a) we show the dependence of the torque
${\cal T}$ on the rotation rate of the outer cylinder in steady
state for both the unlinked actin network and the highest $R$
value used. The (red) line is the best linear fit to the data on
this logarithmic plot demonstrating a power-law dependence of the
measured torque upon the shear rate.

Systems displaying this sort of nonlinear rheology are typically
referred to as power-law fluids~\cite{Barnes:89,Holdsworth:93}, in
which case one has the following relation between stress and rate
of strain:
\begin{equation}
\sigma = \eta' \dot{\gamma}^{n}\propto\Omega^n,
\end{equation}
where $n$ is the exponent of the power-law fluid. In such a
system, the torque versus rotation curve can be used to determine
the exponent in the relation between stress and rate of strain. In
this case, the effective viscosity ($\sigma/\dot{\gamma}$) depends
on shear rate as
\begin{equation}
\label{power-law-fluid}
\eta = \eta' \dot{\gamma}^{n-1},
\end{equation}
As discussed above for a case of a Newtonian fluid, it is possible
to extract the effective viscosity from the measured torque ${\cal
T}$ at the inner cylinder and the imposed angular velocity of the
outer cylinder $\Omega$, once the value of $n$ is determined. In
this case, however, we must modify Eq.~\ref{torque-viscosity} to
account for the nonlinear dependence of stress on shear rate given
by Eq.~\ref{power-law-fluid}.

For a power-law fluid, the relationship between the measured
stress at the inner rotor (determined from
Eq.~\ref{Torque-Stress}) and the angular velocity of the outer
rotor $\Omega$ is given by~\cite{Bird:77}
\begin{equation}
\label{power-viscosity-I} \sigma = \eta \dot{\gamma} = \frac{2
\eta}{n}\frac{\Omega r_o^{2/n}}{r_o^{2/n} - r_i^{2/n}},
\end{equation}
so the effective viscosity can be calculated in terms of the
experimentally measured parameters using
Eqs.~\ref{Torque-Stress} and \ref{power-viscosity-I}. We find that
\begin{equation}
\label{power-viscosity-II} \eta = \frac{{\cal T} (r_o^{2/n} -
r_i^{2/n})n}{4\pi\Omega r_i^{2} r_o^{2/n}}.
\end{equation}
One does have to be concerned that slip at the walls may play a
role in influencing the measured values of the viscosity. Based on
previous work, however, we do not expect slip to play a role in this
system~\cite{Twardos:03}.

In Fig.~\ref{fig:power-law}(b) we extract the effective viscosity
from these data using Eq.~\ref{power-viscosity-II}. The (red) line
again represents the best linear fit to the data on a logarithmic
plot and shows that the effective viscosity indeed has a power-law
dependence on shear rate. The F-actin network is shear thinning
over all shear rates probed encompassing a range of $10^{-2}
\mbox{s}^{-1}$ to $1 \mbox{s}^{-1}$. From these data we find that
the exponent is best fit by  $n = 0.65 \pm 0.02$ for the
uncross-linked sample and $n = 0.52 \pm 0.04$ for the $R = 0.04$.
Notice, the variation in the magnitude of the viscosity from
increasing the cross linking density is small in the range of
shear rates studied, but there appears to be a small, but
measurable, impact on the exponent. This will be discussed in more
detail later.

We vary the cross-linker concentration by increasing the fraction $R$
of biotynlated to non-biotynlated G-actin in the polymerizing
actin solution. As the concentration of cross-linkers increases,
we observe well-defined trends in both the elastic response of the
network at small strains and the steady-state, nonlinear rheology
of the network. We expect that the actual number of cross-linkers
in the network is proportional to the fraction of the these
cross-linking molecules. Clearly, at high enough concentrations of
cross-linkers this linear relationship fails as steric constraints
prevent the saturation of the available biotyn-bonds.

\begin{figure}
\includegraphics[width=6cm]{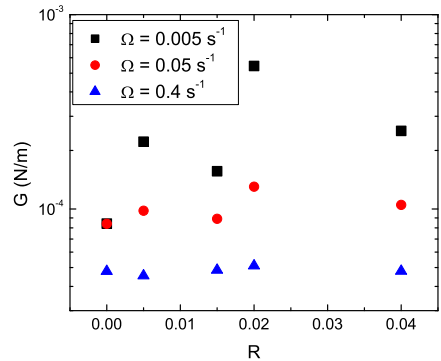}
\caption{(color online) The elastic modulus as extracted from the
data shown in Fig.~\protect{\ref{fig:TorquevsTime}} vs.\ cross-linker density as
measured by $R$.}
\label{fig:ElasticModulusvsCrossLink}
\end{figure}

In Fig.~\ref{fig:ElasticModulusvsCrossLink}, we extract the
elastic modulus Eq.~\ref{elastic-mod-integral} for a range of
cross-linker concentrations. At each concentration we study the
system at three different shear rates. At the lowest shear rate
studied (black squares), we see a roughly linear increase in
elastic modulus with cross-linker concentration. At the highest
concentration for which we took data, the modulus appears to have
reached a plateau, which we attribute to the failure of the system
to make use of the additional cross-linkers on the filaments.
The linear dependence of the elastic modulus is expected in an
affinely deforming~\cite{Head:03,Wilhelm:03} network, and the
eventual saturation is not unexpected for reasons discussed above.
Turning to the data taken at the intermediate shear rate (red
circles) we see that the linear increase of the observed modulus
with cross-linker concentration remains but at a smaller slope.
Finally, at the highest shear rate used (blue triangles), there is
no observable dependence of the modulus on the cross-linker
concentration.

\begin{figure}
\includegraphics[width=6cm]{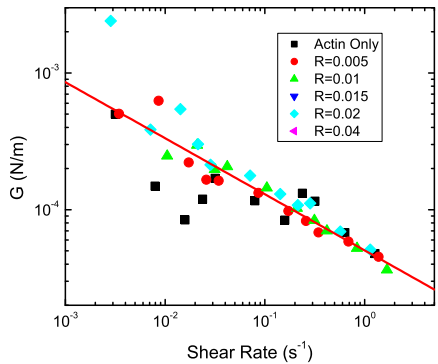}
\caption{(color online) The elastic modulus vs. rotation rate of
the outer cylinder for a range of cross-linker concentrations. The
linear dependence (best fit line in red) demonstrates a power law
dependence of the elastic modulus on rotation rate with an
exponent of $-0.4$.} \label{fig:ElasticModulus-shear-rate}
\end{figure}

We also note from Fig.~\ref{fig:ElasticModulusvsCrossLink} that
the magnitude of the elastic response of the network extracted
from the initiation of shear decreases with increasing shear rate.
This is illustrated in more detail in
Fig.~\ref{fig:ElasticModulus-shear-rate} The measured modulus
decreases with imposed shear rate in a manner consistent with a
power-law
\begin{equation}
\label{G-shear-rate-dependence} G'(\dot{\gamma}) \sim
\dot{\gamma}^{-0.4}.
\end{equation}
The data corresponding to higher shear rates is essentially
independent of cross-link density. We interpret this behavior as
suggesting that the observed modulus in this high shear rate
regime is due to the mechanics of interacting domains of
cross-linked filaments, and not elastic strain energy stored in
individual cross-linked networks.

\begin{figure}
\includegraphics[width=6cm]{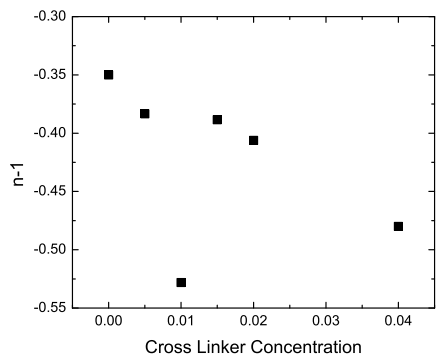}
\caption{The dependence of the best fit value of
$n-1$ for the dependence of the viscosity on rate of strain as a
function of ratio of cross-linker to G-actin concentration.}
\label{fig:VSRvsCrossLinker}
\end{figure}

For comparison, it is worth returning to the impact of the
cross-link density on the viscosity. What is interesting is that
even though the density of cross-linkers does not appear to impact
the magnitude of the measured viscosity significantly, there is a
weak effect on the exponent $n$ from the fits to a power-law
fluid. In Fig.~\ref{fig:VSRvsCrossLinker}, we plot the exponent
for the viscosity ($n - 1$), where $n$ is defined in
Eq.~\ref{power-law-fluid}. As the cross-linker concentration
increases the exponent $n$ decreases from $n=0.65 \pm 0.02$ to
$n=0.52 \pm 0.04$. Unlike the case of the elastic modulus, we do
not find any saturation of the exponent with increasing
cross-linker concentration.

Sollich et al.\ have proposed~\cite{Sollich:97} this sort of
power-law rheology can arise rather generically in the mechanics
of structurally disordered, soft materials that exhibit at large
number of metastable microstates consistent with a given applied
stress. Given our microscopic picture of long F-actin filaments
slipping past each other under the applied shear stress, this
unifying picture of the rheology  of soft
disordered materials appears applicable to the strongly sheared
network. Due to steric interactions between the filaments we
expect to find such rheological behavior in both the cross-linked
(by biotyn-streptavidin) and the uncross-linked samples.

In the Sollich framework, we may interpret the power law exponent
$n$ in terms of as ``effective noise temperature'' $x = n + 1$.
Over the range of cross-linker concentrations studied $x$ ranges
from $x = 1.35$ to $x = 1.45$. Given these values of the effective
noise temperature, the system is predicted to exhibit power-law
rheology in steady-state. Moreover, they predict that the system
should have a vanishing yield stress. In other words it should
flow in the limit of vanishingly small applied shears. While we
have focused on the steady-flow behavior of the F-actin/monolayer
material, our preliminary low shear stress data are are consistent
with this prediction. We will report more fully on the extremely
low shear regime elsewhere~\cite{Walder:07a}.

\section{Conclusions}
\label{conclusions}

The complex mechanics of cross-linked F-actin networks underlies
the dynamics and mechanics of living cells. While there has been a
great deal of recent work on development of {\em in vitro}
semiflexible network models of the cytoskeleton and measurements
of their rheology, comparatively little effort has been expended
on the development of analogous model systems to explore the
mechanical interaction of the cytoskeleton with the plasma
membrane. Here we have reported on just such a simplified model
system that incorporates a mechanical coupling between a
cross-linked semiflexible network and a lipid monolayer that
mimics one leaflet of the plasma membrane of the cell. This model
is clearly a minimal description of cytoskeletal networks
mechanically associated with a liquid lipid phase boundary.

We have characterized the steady-state shear response of this composite
material and found it to be well-described at higher shear rates
by a power-law fluid model with an exponent consistent with those
obtained from the soft glassy rheology picture of soft disordered materials.
Such a correspondence is not entirely unexpected since, at the filament
densities studied we imagine the system to consist of cross-linked
and sterically confined long filaments that have numerous mechanical
equilibria associated with moving uncross-linked strands of the network
around each other.

We find at higher shear rates the effective viscosity decreases as
$\eta \sim \dot{\gamma}^{-0.4}$ and the elastic modulus decreases
as $G' \sim \dot{\gamma}^{-0.4}$. The impact of the cross-linker
density is interesting. At lower shear rates the elastic modulus
depends roughly linearly upon cross-linker concentration although
at higher shear rates we find essentially no rheological effect of
the cross-linker concentration -- see Fig.~\ref{fig:ElasticModulusvsCrossLink}.
The relaxation time after the
cessation of flow
is independent of the cross-linker density. The
magnitude of the viscosity does not depend in any systematic way
on cross-linker density, but the power-law used to describe the
steady state viscosity decreases as a function of cross-linker
density as shown in Fig.~\ref{fig:VSRvsCrossLinker}.
Combining these facts strongly suggests the following
picture. The system is composed of clusters of cross-linked
filaments. Our results implies that the observed mechanics are due
to cross-linked filaments for low shear rates and strains (the
elastic modulus), but at higher shear rates and strains, we are
probing the interaction between clusters of cross-linker filaments
(viscosity measurements, relaxation time after shear, and high
shear rate elastic modulus).

>From this picture, one is not surprised that the soft-glassy
rheology ideas are appropriate for describing the steady state
behavior, as the clusters will act as the fundamental mechanical
elements in the soft glassy system. However, the dependence on the
power-law exponent on cross-linker density provides an interesting
direction for further study. This suggests that the details of the
cluster-cluster interaction depend on the intrinsic mechanical
properties of the clusters themselves. These are clearly dependent
on the cross-linker density, as indicated by the low shear rate
measurements of the elastic modulus. This presents us with an
interesting system to study the impact of local mechanical
properties on the macroscopic properties. A detailed study of the
internal cluster properties will be the subject of future work
using microrheological probes of the network associated with the
lipid monolayer.

%final comment in response to Referee A

Microrheological studies will also be important as future studies
incorporate additional complexity of linkers. For example, when one
considers linkers that modify the local structure of the network (e.g.
cause bundling) it will be important to have simultaneous access to
both the local elastic response of the network and the system-size,
macroscopic elasticity of the gel. Additionally,
a combination of traditional (Couette) rheometry and microrheological studies
will be helpful in accessing the rheological effect of labile, physiological
cross-linkers.  It is clear that the low-frequency linear response
of a network cross-linked
by such transient bonds will be
dramatically different from our biotin/streptavidin cross-linked system. In our case
the network is a soft solid, having a finite elastic response at zero frequency. The
network with labile cross-linkers, however,  must flow like a liquid on time scales longer than
the typical binding lifetime of a transient cross-linker. To probe this low-frequency
rheology, the traditional Couette cell approach can be used. The addition of
microrheology allows one to explore the higher frequency response of the system that
is inaccessible to the macroscopic measurements. These high frequency data should
show that the presence of labile cross-linkers affect only the low frequency
response of the system. The effect of labile cross-linkers on the nonlinear rheology
of the system at high strains also remains to be considered.

\section{Acknowledgements}
We acknowledge the support of NSF-DMR-0354113 and Christoph
Schmidt for helpful discussions. R. Walder acknowledges support
through a travel fellowship from the Institute for Complex
Adaptive Matter.

\end{document}